\documentclass[12pt,a4paper]{article}
\makeatletter
\def\eqnarray{%
\stepcounter{equation}%
\let\@currentlabel=\theequation
\global\@eqnswtrue
\global\@eqcnt\z@
\tabskip\@centering
\let\\=\@eqncr
$$\halign to \displaywidth\bgroup\@eqnsel\hskip\@centering
$\displaystyle\tabskip\z@{##}$&\global\@eqcnt\@ne
\hfil$\displaystyle{{}##{}}$\hfil
&\global\@eqcnt\tw@$\displaystyle\tabskip\z@{##}$\hfil
\tabskip\@centering&\llap{##}\tabskip\z@\cr}
\makeatother
\newcommand{\kansu}[2]{{{#1}\!\left({#2}\right)}}
\newcommand{\ket}[1]{{\vert{#1}\rangle}}
\newcommand{\bra}[1]{{\langle{#1}\vert}}
\newcommand{\kett}[2]{{\vert{#1,#2}\rangle}}
\newcommand{\braa}[2]{{\langle{#1,#2}\vert}}
\newcommand{\braket}[2]{{\langle{#1}\vert{#2}\rangle}}
\newcommand{\zettai}[1]{{\vert{#1}\vert}}

\newcommand{\calh}{{\cal H}}
\newcommand{\calm}{{\cal M}}

\newcommand{\fukuso}{{\mathbf C}}

\newcommand{\futon}{{\bf N}}

\newcommand{\stm}{{St_m}}
\newcommand{\grm}{{Gr_m}}
\newcommand{\eem}{{E_m}}

\newcommand{\szetta}[1]{{\vert{#1}\vert}^{2}}

\newcommand{\zetta}{{\vert z\vert}}
\newcommand{\wetta}{{\vert w\vert}}

\begin{document}

\title{\sl Geometry of Generalized Coherent States : Some 
           Calculations of Chern Characters}
\author{
  Kazuyuki FUJII
  \thanks{E-mail address : fujii@math.yokohama-cu.ac.jp}\  
  \thanks{Home-page : http://fujii.sci.yokohama-cu.ac.jp}\\
  Department of Mathematical Sciences\\
  Yokohama City University\\
  Yokohama 236-0027\\ 
  JAPAN
  }
\date{}
\maketitle\thispagestyle{empty}
%
%
%  gaiyou
%
%
\begin{abstract}
  This is a continuation of the preceding paper (hep--ph/0108219). 
 
  First of all we make a brief review of generalized coherent states 
  based on Lie algebra su(1,1) and prove that the resolution of unity 
  can be obtained by the curvature form of some bundle.   
  
  Next for a set of generalized coherent states we define a universal 
  bundle over the infinite--dimensional Grassmann manifold and construct 
  the pull--back bundle making use of a projector from the parameter space 
  to this Grassmann one. 
  We mainly study Chern characters of these bundles. 

  Although the Chern characters in the infinite--dimensional case 
  are in general not easy to calculate, we can perform them for the 
  special cases. In this paper we report our calculations and propose 
  some problems. 
\end{abstract}

\newpage

%
%
%     Honbun
%
%

\section{Introduction}

Coherent states or generalized coherent states play very important role 
in quantum physics, in particular, quantum optics, see \cite{KS} and  
its references or \cite{MW}.  
They also play an important one in mathematical physics. See 
the book \cite{AP}. For example, they are very useful in performing 
stationary phase approximations to path integral, see \cite{FKSF1}, 
\cite{FKSF2} and \cite{FKS}. 

This is a continuation of the preceding paper \cite{KF9}.  
Namely, since we have studied coherent states from the geometric point 
of view, we continue to study generalized coherent states from the 
same point of view. For a set of generalized coherent states we can define 
a projector from the manifold consisting of parameters of them to 
infinite--dimensional Grassmann manifold.  
Making use of this we can calculate several geometric quantities, see 
for example \cite{MN}. 

In this paper we mainly focus on Chern characters because they play an 
very important role in global geometry. But their calculations are not 
so easy. Our calculations are only $m=1,\ 2$ (see the section 3). Even 
the case $m=2$ the calculations are very complicated. We must leave 
the case $m=3$ to the readers with high powers. 

The hidden aim of this paper and \cite{KF9} is to apply 
the results to Quantum Information Theory (QIT) including Quantum 
Computation (QC) $\cdots$ {\bf a geometric construction of QIT}.  
As for QC or QIT see \cite{LPS}, \cite{AH} and \cite{KF1} for 
general introduction. 
We are in particular interested in Holonomic Quantum Computation, see 
\cite{KF2}--\cite{KF6}. We are also interested in Homodyne Tomography 
\cite{GLM}, \cite{MP} or Quantum Cryptgraphy \cite{KB}, \cite{BW}. 
In sequel papers we will discuss these points. 

By the way it seems to the author that our calculations suggest 
some deep relation to recent non--commutative differential geometry or 
non--commutative field theory, see \cite{SV} or \cite{BDLMO}. 
But this is beyond the scope of this paper. 

The author expects strongly that young mathematical physicists or 
information theorists will enter to this fruitful field.

\vspace{10mm}
\section{Coherent States and Generalized Coherent Ones}

\subsection{Coherent States}

We make a brief review of some basic properties of coherent operators 
within our necessity, \cite{KS} and \cite{AP}.

Let $a(a^\dagger)$ be the annihilation (creation) operator of the harmonic 
oscillator.
If we set $N\equiv a^\dagger a$ (:\ number operator), then
\begin{equation}
  \label{eq:2-1}
  [N,a^\dagger]=a^\dagger\ ,\
  [N,a]=-a\ ,\
  [a^\dagger, a]=-\mathbf{1}\ .
\end{equation}
Let $\calh$ be a Fock space generated by $a$ and $a^\dagger$, and
$\{\ket{n}\vert\  n\in\futon\cup\{0\}\}$ be its basis.
The actions of $a$ and $a^\dagger$ on $\calh$ are given by
\begin{equation}
  \label{eq:2-2}
  a\ket{n} = \sqrt{n}\ket{n-1}\ ,\
  a^{\dagger}\ket{n} = \sqrt{n+1}\ket{n+1}\ ,
  N\ket{n} = n\ket{n}
\end{equation}
where $\ket{0}$ is a normalized vacuum ($a\ket{0}=0\  {\rm and}\  
\langle{0}\vert{0}\rangle = 1$). From (\ref{eq:2-2})
state $\ket{n}$ for $n \geq 1$ are given by
\begin{equation}
  \label{eq:2-3}
  \ket{n} = \frac{(a^{\dagger})^{n}}{\sqrt{n!}}\ket{0}\ .
\end{equation}
These states satisfy the orthogonality and completeness conditions
\begin{equation}
  \label{eq:2-4}
   \langle{m}\vert{n}\rangle = \delta_{mn}\ ,\quad \sum_{n=0}^{\infty}
   \ket{n}\bra{n} = \mathbf{1}\ . 
\end{equation}

 Let us state coherent states. For the normalized state $\ket{z} \in 
\calh \ {\rm for}\  z \in \fukuso$ the following three conditions are 
equivalent :
\begin{eqnarray}
  \label{eq:2-5-1}
 &&(\mbox{i})\quad a\ket{z} =  z\ket{z}\quad {\rm and}\quad 
      \langle{z}\vert{z}\rangle = 1  \\
  \label{eq:2-5-2}
 &&(\mbox{ii})\quad  \ket{z} =  \mbox{e}^{- \vert{z}\vert^{2}/2} 
          \sum_{n=0}^{\infty}\frac{z^{n}}{\sqrt{n!}}\ket{n} = 
          \mbox{e}^{- \vert{z}\vert^{2}/2}e^{za^{\dagger}}\ket{0} \\
  \label{eq:2-5-3}
 &&(\mbox{iii})\quad  \ket{z} =  \mbox{e}^{za^{\dagger}- \bar{z}a}\ket{0}. 
\end{eqnarray}
In the process from (\ref{eq:2-5-2}) to (\ref{eq:2-5-3}) 
we use the famous elementary Baker-Campbell-Hausdorff formula
\begin{equation}
  \label{eq:2-6}
 \mbox{e}^{A+B}=\mbox{e}^{-\frac1{2}[A,B]}\mbox{e}^{A}\mbox{e}^{B}
\end{equation}
whenever $[A,[A,B]] = [B,[A,B]] = 0$, see \cite{KS} or \cite{AP}.  
This is the key formula.

\noindent{\bfseries Definition}\quad The state $\ket{z}$ that 
satisfies one of (i) or (ii) or (iii) above is called the coherent state.

\noindent
The important feature of coherent states is the following partition 
(resolution) of unity.
\begin{equation}
  \label{eq:2-7}
  \int_{\fukuso} \frac{[d^{2}z]}{\pi} \ket{z}\bra{z} = 
  \sum_{n=0}^{\infty} \ket{n}\bra{n} = \mathbf{1}\ ,
\end{equation}
where we have put $[d^{2}z] = d(\mbox{Re} z)d(\mbox{Im} z)$ for simplicity. 
We note that 
\begin{equation}
   \label{eq:2-a}
  \braket{z}{w} = \mbox{e}^{-\frac{1}{2}\zetta^2 -\frac{1}{2}\wetta^2 +
                             \bar{z}w} \Longrightarrow 
  \vert{\braket{z}{w}}\vert=
  \mbox{e}^{-\frac{1}{2}\vert{z-w}\vert^2},\  
  \braket{w}{z}=\overline{\braket{z}{w}},
\end{equation}
so $\vert{\braket{z}{w}}\vert < 1$ if $z\ne w$ and 
$\vert{\braket{z}{w}}\vert \ll 1$ if $z$ and $w$ are separated enough. 
We will use this fact in the following. 

Since the operator
\begin{equation}
  \label{eq:2-8}
      U(z) = \mbox{e}^{za^{\dagger}- \bar{z}a}
      \quad \mbox{for} \quad z \in \fukuso  
\end{equation}
is unitary, we call this a (unitary) coherent operator. For these 
operators the following properties are crucial.
For  $z,\ w \in \fukuso$
\begin{eqnarray}  
  \label{eq:2-9-1} 
 && U(z)U(w) = \mbox{e}^{z\bar{w}-\bar{z}w}\ U(w)U(z), \\
  \label{eq:2-9-2}
 && U(z+w) = \mbox{e}^{-\frac{1}{2}(z\bar{w}-\bar{z}w)}\ U(z)U(w).
\end{eqnarray}

\subsection{Generalized Coherent States}

Next we make a brief review of some basic properties of generalized 
coherent operators based on $su(1,1)$, see \cite{FKSF1} or \cite{AP}.     

We consider a spin $K\ (> 0)$ representation of $su(1,1) 
\subset sl(2,\fukuso)$ and set its generators 
$\{ K_+, K_-, K_3 \}\ ((K_+)^{\dagger} = K_-)$, 
\begin{equation}
  \label{eq:2-2-12}
 [K_3, K_+]=K_+, \quad [K_3, K_-]=-K_-, \quad [K_+, K_-]=-2K_3.
\end{equation}
We note that this (unitary) representation is necessarily infinite 
dimensional. 
The Fock space on which $\{ K_+, K_-, K_3 \}$ act is 
$\calh_{K} \equiv \{\kett{K}{n} \vert n\in\futon\cup\{0\} \}$ and 
whose actions are
\begin{eqnarray}
  \label{eq:2-2-13}
 K_+ \kett{K}{n} &=& \sqrt{(n+1)(2K+n)}\kett{K}{n+1} , \nonumber \\
 K_- \kett{K}{n} &=& \sqrt{n(2K+n-1)}\kett{K}{n-1} ,  \nonumber   \\
 K_3 \kett{K}{n} &=& (K+n)\kett{K}{n}, 
\end{eqnarray}
where $\kett{K}{0}$ is a normalized vacuum ($K_-\kett{K}{0}=0$ and 
$\langle K,0|K,0 \rangle =1$). We have written $\kett{K}{0}$ instead 
of $\ket{0}$  to emphasize the spin $K$ representation, see \cite{FKSF1}. 
We also denote by ${\bf 1}_{K}$ the unit operator on $\calh_K$. 
From (\ref{eq:2-2-13}), states $\kett{K}{n}$ are given by
\begin{equation}
  \label{eq:2-2-14}
 \kett{K}{n} =\frac{(K_+)^n}{\sqrt{n!(2K)_n}}\kett{K}{0} ,
\end{equation}
where $(a)_n$ is the Pochammer's notation 
$(a)_n \equiv  a(a+1) \cdots (a+n-1)$. 
These states satisfy the orthogonality and completeness conditions 
\begin{equation}
  \label{eq:2-2-16}
  \langle K,m \vert K,n \rangle =\delta_{mn}, 
 \quad \sum_{n=0}^{\infty}\kett{K}{n}\braa{K}{n}\ = \mathbf{1}_K.
\end{equation}
Now let us consider a generalized version of coherent states : 

\noindent{\bfseries Definition}\quad We call a state 
\begin{equation}
   \label{eq:2-2-17}
 \ket{w}  \equiv \mbox{e}^{wK_+ - \bar{w}K_-} \kett{K}{0}  
  \quad \mbox{for} \quad w \in \fukuso.
\end{equation}
the generalized coherent state (or the coherent state of Perelomov's 
type based on $su(1,1)$ in our terminology, \cite{FS}).

\noindent  
We note that this is the extension of (\ref{eq:2-5-3}) not (\ref{eq:2-5-1}), 
see \cite{AP}. 
For this the following disentangling formula is well--known : 
\begin{eqnarray}
  \label{eq:2-2-20}
    \mbox{e}^{wK_{+} -\bar{w}K_{-}}  
  &=& \mbox{e}^{\zeta K_+}\mbox{e}^{\log (1-\vert\zeta\vert^2)K_3}
    \mbox{e}^{-\bar{\zeta}K_-} \quad \mbox{or}   \nonumber \\
  &=& \mbox{e}^{-\bar{\zeta}K_-}\mbox{e}^{-\log (1-\vert\zeta\vert^2)K_3}
    \mbox{e}^{\zeta K_+}.
\end{eqnarray}
where 
\begin{equation}
    \zeta = \zeta(w) \equiv 
    \frac{{w} \mbox{tanh}(\zettai{w})}{\zettai{w}}\ \ 
   (\Longrightarrow \zettai{\zeta} < 1). 
\end{equation}
This is the key formula for generalized coherent operators. Therefore from 
(\ref{eq:2-2-13}) 
\begin{equation}
  \label{eq:2-2-right}  
   \ket{w}=(1-\zettai{\zeta}^2)^{K}\mbox{e}^{\zeta K_{+}}\kett{K}{0}
          \equiv \ket{\zeta}.
\end{equation}
This corresponds to the right hand side of (\ref{eq:2-5-2}). Moreover 
since 
\[
    \mbox{e}^{\zeta K_{+}}\kett{K}{0}
      = \sum_{n=0}^{\infty}\frac{\zeta^n}{n!}K_{+}^{n}\kett{K}{0}
      = \sum_{n=0}^{\infty}\sqrt{\frac{(2K)_n}{n!}}
        \frac{\zeta^n K_{+}^{n}}{\sqrt{(2K)_{n}n!}}\kett{K}{0}
      = \sum_{n=0}^{\infty}\sqrt{\frac{(2K)_n}{n!}}\zeta^n \kett{K}{n}
\]
we have 
\begin{equation}
  \label{eq:2-2-left}  
   \ket{w}=(1-\zettai{\zeta}^2)^{K}
    \sum_{n=0}^{\infty}\sqrt{\frac{(2K)_n}{n!}}\zeta^n \kett{K}{n}
   \equiv \ket{\zeta}.
\end{equation}
This corresponds to the left hand side of (\ref{eq:2-5-2}).  
Then the partition of unity corresponding to (\ref{eq:2-7}) is 
\begin{eqnarray}
  \label{eq:2-2-18}  
    &&\int_{\fukuso}\frac{2K-1}{\pi} \frac{\mbox{tanh}(\zettai{w})[d^{2}w]}
     {\left(1-\mbox{tanh}^2(\zettai{w})\right)\zettai{w}}
     \ket{w}\bra{w}  \nonumber \\
   = &&\int_{\mbox{D}}\frac{2K-1}{\pi} \frac{[d^{2}\zeta]}{\left(1- \vert
    \zeta\vert^{2}\right)^2} \ket{\zeta}\bra{\zeta} = 
    \sum_{n=0}^{\infty}\kett{K}{n}\braa{K}{n}\ = \mathbf{1}_K,
\end{eqnarray}
where 
$\fukuso \rightarrow \mbox{D} : w \mapsto \zeta = \zeta(w) $ and 
$D$ is the Poincare disk in $\fukuso$, see \cite{KF7}. 

Here let us construct the spin $K$--representations making 
use of Schwinger's boson method.

\par \noindent 
First we set 
\begin{equation}
  \label{eq:2-2-21}
  K_+\equiv{1\over2}\left( a^{\dagger}\right)^2\ ,\
  K_-\equiv{1\over2}a^2\ ,\
  K_3\equiv{1\over2}\left( a^{\dagger}a+{1\over2}\right)\ , 
\end{equation}
then we have
\begin{equation}
  \label{eq:2-2-22}
  [K_3,K_+]=K_+\ ,\
  [K_3,K_-]=-K_-\ ,\
  [K_+,K_-]=-2K_3\ .
\end{equation}
That is, the set $\{K_+,K_-,K_3\}$ gives a unitary representation of $su(1,1)$
with spin $K = 1/4\ \mbox{and}\ 3/4$. Now we also call an operator 
\begin{equation}
  \label{eq:2-2-23}
   S(w) = \mbox{e}^{\frac{1}{2}\{w(a^{\dagger})^2 - \bar{w}a^2\}}
   \quad \mbox{for} \quad w \in \fukuso 
\end{equation}
the squeezed operator.

Next we consider the system of two-harmonic oscillators. If we set
\begin{equation}
  \label{eq:2-2-24}
  a_1 = a \otimes 1,\  {a_1}^{\dagger} = a^{\dagger} \otimes 1;\ 
  a_2 = 1 \otimes a,\  {a_2}^{\dagger} = 1 \otimes a^{\dagger},
\end{equation}
then it is easy to see 
\begin{equation}
  \label{eq:2-2-25}
 [a_i, a_j] = [{a_i}^{\dagger}, {a_j}^{\dagger}] = 0,\ 
 [a_i, {a_j}^{\dagger}] = \delta_{ij}, \quad i, j = 1, 2. 
\end{equation}
We also denote by $N_{i} = {a_i}^{\dagger}a_i$ number operators.

Now we can construct representation of Lie algebras $su(2)$ and $su(1,1)$ 
making use of Schwinger's boson method, see \cite{FKSF1}, \cite{FKSF2}. 
Namely if we set
\begin{eqnarray}
  \label{eq:2-2-26-1}
  su(2) &:&\quad
     J_+ = {a_1}^{\dagger}a_2,\ J_- = {a_2}^{\dagger}a_1,\ 
     J_3 = {1\over2}\left({a_1}^{\dagger}a_1 - {a_2}^{\dagger}a_2\right), \\
  \label{eq:2-2-26-2}
  su(1,1) &:&\quad
     K_+ = {a_1}^{\dagger}{a_2}^{\dagger},\ K_- = a_2 a_1,\ 
     K_3 = {1\over2}\left({a_1}^{\dagger}a_1 + {a_2}^{\dagger}a_2  + 1\right),
\end{eqnarray}
then we have
\begin{eqnarray}
  \label{eq:2-2-27-1}
  su(2) &:&\quad
     [J_3, J_+] = J_+,\ [J_3, J_-] = - J_-,\ [J_+, J_-] = 2J_3, \\
  \label{eq:2-2-27-2}
  su(1,1) &:&\quad
     [K_3, K_+] = K_+,\ [K_3, K_-] = - K_-,\ [K_+, K_-] = -2K_3.
\end{eqnarray}

In the following we define (unitary) generalized coherent operators 
based on Lie algebras $su(2)$ and $su(1,1)$. 

\noindent{\bfseries Definition}\quad We set 
\begin{eqnarray}
  \label{eq:2-2-28-1}
  su(2) &:&\quad 
W(w) = e^{wJ_{+} - \bar{w}J_{-}}\quad {\rm for}\quad w \in \fukuso , \\
  \label{eq:2-2-28-2}
  su(1,1) &:&\quad 
V(w) = e^{wK_{+} - \bar{w}K_{-}}\quad {\rm for}\quad w \in \fukuso.
\end{eqnarray}

\par \noindent 
For the latter convenience let us list well-known disentangling formulas  
once more. We have
\begin{eqnarray}
  \label{eq:j-formula}
  su(2) &:&\quad 
W(w) = e^{\eta J_{+}}
 e^{{\rm log}\left(1 + \zettai{\eta}^{2}\right)J_{3}}
 e^{- \bar{\eta}J_{-}}, \quad 
   where\quad  \eta = \frac{{w}{\rm tan}{(\zettai{w})}}
                           {\zettai{w}},  \\
  \label{eq:k-formula}
  su(1,1) &:&\quad 
V(w) =  e^{\zeta K_{+}}
 e^{{\rm log}\left(1 - \zettai{\zeta}^{2}\right)K_{3}}
 e^{- \bar{\zeta}K_{-}},\quad 
   where\quad  \zeta = \frac{{w}{\rm tanh}{(\zettai{w})}}
                             {\zettai{w}}. 
 \end{eqnarray}
As for a generalization of these formulas see \cite{FS}.\quad  
Before ending this section let us ask a question.

What is a relation between (\ref{eq:2-2-28-2}) and (\ref{eq:2-2-23}) 
of generalized coherent operators based on $su(1.1)$ ?

\noindent
The answer is given by Paris \cite{MP} :

\noindent{\bfseries Formula}\quad We have 
\begin{equation}
  \label{eq:2-2-29}
  W(-\frac{\pi}{4})S_{1}(w)S_{2}(-w) W(-\frac{\pi}{4})^{-1} = V(w),
\end{equation}
where $S_{j}(w)=(\ref{eq:2-2-23})$ with $a_{j}$ instead of $a$. 

\noindent 
Namely, $V(w)$ is given by ``rotating'' the product $S_{1}(w)S_{2}(-w)$ 
by $ W(-\frac{\pi}{4})$. This is an interesting relation.

\subsection{Some Formulas on Generalized Coherent States}

We make some preliminaries for the following section. For that we list 
some useful formulas on generalized coherent states. Since the proofs are 
not so difficult, we leave them to the readers. 

\noindent{\bfseries Formulas}\quad For $w_1, w_2$ we have
\begin{eqnarray}
  \label{eq:2-3-a-1}
 &&(\mbox{i})\quad \braket{w_1}{w_2}=
         \left\{\frac{(1-\zettai{\zeta_1}^2)(1-\zettai{\zeta_2}^2)}
                     {(1-{\bar \zeta_1}\zeta_2)^2} 
         \right\}^{K},    \\
&&{}  \nonumber \\
  \label{eq:2-3-b-2}
 &&(\mbox{ii})\quad 
      \bra{w_1}K_{+}\ket{w_2}=\braket{w_1}{w_2}
                   \frac{2K{\bar \zeta_1}}{1-{\bar \zeta_1}\zeta_2}\ ,  \\
&&{}  \nonumber \\
  \label{eq:2-3-c-3}
 &&(\mbox{iii})\quad  
      \bra{w_1}K_{-}\ket{w_2}=\braket{w_1}{w_2}
                   \frac{2K\zeta_2}{1-{\bar \zeta_1}\zeta_2}\ ,  \\
&&{}  \nonumber \\
  \label{eq:2-3-d-4}
 &&(\mbox{iv})\quad  
      \bra{w_1}K_{-}K_{+}\ket{w_2}=\braket{w_1}{w_2}
         \frac{2K+4K^2{\bar \zeta_1}\zeta_2}{(1-{\bar \zeta_1}\zeta_2)^2}\ .
\end{eqnarray}
where 
\begin{equation}
  \label{eq:two-zetas}
     \zeta_j=\frac{{w_j}\mbox{tanh}(\zettai{w_j})}{\zettai{w_j}}\quad 
             \mbox{for}\quad j=1,\ 2.
\end{equation}
When $w_1=w_2\equiv w$, then $\braket{w}{w}=1$, so we have 
\begin{eqnarray}
  \label{eq:kitaichi-1}
      &&\bra{w}K_{+}\ket{w}=
             \frac{2K{\bar \zeta}}{1-\zettai{\zeta}^2}\ , \quad 
      \bra{w}K_{-}\ket{w}
             \frac{2K\zeta}{1-\zettai{\zeta}^2}\ , \\
  \label{eq:kitaichi-2}
      &&\bra{w}K_{-}K_{+}\ket{w}=
         \frac{2K+4K^2\zettai{\zeta}^2}{(1-\zettai{\zeta}^2)^2}\ .
\end{eqnarray}

\par \noindent 
Here let us make a comment. From (\ref{eq:2-3-a-1}) 
\[
  \zettai{\braket{w_1}{w_2}}^2=
       \left\{\frac{(1-\zettai{\zeta_1}^2)(1-\zettai{\zeta_2}^2)}
                   {\zettai{1-{\bar \zeta_1}\zeta_2}^2} 
       \right\}^{2K}, 
\]
so we want to know the property of 
\[
   \frac{(1-\zettai{\zeta_1}^2)(1-\zettai{\zeta_2}^2)}
        {\zettai{1-{\bar \zeta_1}\zeta_2}^2}. 
\]
It is easy to see that
\begin{equation}
 \label{eq:futou-shiki-1}
 1 - \frac{(1-\zettai{\zeta_1}^2)(1-\zettai{\zeta_2}^2)}
          {\zettai{1-{\bar \zeta_1}\zeta_2}^2}
 =\frac{\zettai{\zeta_1-\zeta_2}^2}{\zettai{1-{\bar \zeta_1}\zeta_2}^2}
 \geq 0
\end{equation}
and $(\ref{eq:futou-shiki-1})=0$ if and only if (iff) $\zeta_1=\zeta_2$. 
Therefore 
\begin{equation}
 \label{eq:futou-shiki-2}
 \zettai{\braket{w_1}{w_2}}^2 = 
       \left\{\frac{(1-\zettai{\zeta_1}^2)(1-\zettai{\zeta_2}^2)}
                   {\zettai{1-{\bar \zeta_1}\zeta_2}^2} 
       \right\}^{2K} \leq 1
\end{equation}
because $2K>1$ ($2K-1 > 0$ from (\ref{eq:2-2-18})). Of course 
\begin{equation}
 \label{eq:futou-shiki-3}
  \zettai{\braket{w_1}{w_2}}=1\quad \mbox{iff}\quad \zeta_1=\zeta_2 
                              \quad \mbox{iff}\quad  w_1=w_2 .
\end{equation}
by (\ref{eq:two-zetas}).

\subsection{A Supplement on Generalized Coherent States}

Before ending this section let us make a brief comment on generalized 
coherent states (\ref{eq:2-2-17}). The coherent states $\ket{z}$ has been 
defined by (\ref{eq:2-5-1}) : $a\ket{z}=z\ket{z}$. Why do we define 
the generalized coherent states $\ket{w}$ as $K_{-}\ket{w}=w\ket{w}$ 
because $K_{-}$ is an annihilation operator corresponding to $a$ ? 
First let us try to calculate $K_{-}\ket{w}$ making use of 
(\ref{eq:2-2-right}). 
\[
  K_{-}\ket{w}=
     (1-\zettai{\zeta}^2)^{K}K_{-}\mbox{e}^{\zeta K_{+}}\kett{K}{0}
  =(1-\zettai{\zeta}^2)^{K}\mbox{e}^{\zeta K_{+}}\mbox{e}^{-\zeta K_{+}}
    K_{-}\mbox{e}^{\zeta K_{+}}\kett{K}{0}.
\]
Here it is easy to see 
\begin{eqnarray}
  \mbox{e}^{-\zeta K_{+}}K_{-}\mbox{e}^{\zeta K_{+}}
  &=&\sum_{n=0}^{\infty}\frac{1}{n!}
    [-\zeta K_{+},[-\zeta K_{+},[\ , \cdots, [-\zeta K_{+},K_{-}]\cdots ] ]]
     \nonumber \\
  &=&K_{-}+2\zeta K_{3}+\zeta^{2}K_{+}\ , \nonumber
\end{eqnarray}
from the relations (\ref{eq:2-2-12}), so that 
\begin{eqnarray}
   K_{-}\ket{w}&=&(1-\zettai{\zeta}^2)^{K}\mbox{e}^{\zeta K_{+}}
          (K_{-}+2\zeta K_{3}+\zeta^{2}K_{+})\kett{K}{0} \nonumber \\
   &=&2\zeta K(1-\zettai{\zeta}^2)^{K}\mbox{e}^{\zeta K_{+}}\kett{K}{0}
    + \zeta^{2}K_{+}(1-\zettai{\zeta}^2)^{K}\mbox{e}^{\zeta K_{+}}\kett{K}{0}
    \nonumber \\
   &=&(2K\zeta +\zeta^{2}K_{+})\ket{w} 
\end{eqnarray}
because $K_{-}\kett{K}{0}={\bf 0}$. Namely we have the equation
\begin{equation}
  \label{eq:perolomov-1}
 (K_{-}-\zeta^{2}K_{+})\ket{w}=2K\zeta \ket{w},\quad \mbox{where}\quad 
    \zeta=\frac{{w} \mbox{tanh}(\zettai{w})}{\zettai{w}}, 
\end{equation}
or more symmetrically 
\begin{equation}
  \label{eq:perolomov-2}
 ({\zeta}^{-1}K_{-}-\zeta K_{+})\ket{w}=2K\ket{w},\quad \mbox{where}\quad 
    \zeta=\frac{{w} \mbox{tanh}(\zettai{w})}{\zettai{w}}.  
\end{equation}
This equation is completely different from (\ref{eq:2-5-1}). 

\par \noindent
{\bf A comment is in order}. The states $\vert\vert{w}\rangle\rangle$ 
($w \in \fukuso$) defined by 
\begin{equation}
     K_{-}\vert\vert{w}\rangle\rangle=w\vert\vert{w}\rangle\rangle 
\end{equation}
are called the Barut--Girardello coherent states, \cite{BGi}. The 
solution is given by 
\begin{equation}
  \vert\vert{w}\rangle\rangle=\sum_{n=0}^{\infty}
     \frac{w^n}{\sqrt{n!(2K)_n}}\kett{K}{n}
\end{equation}
up to the normalization factor. Compare this with (\ref{eq:2-2-left}). 

\par \noindent 
Their states have several interesting structures, but we don't consider 
them in this paper.  
See \cite{FF1}, \cite{FF2} and \cite{Tri} as to further developments and 
applications.

\vspace{10mm}
\section{Universal Bundles and Chern Characters}

We make a brief review of some basic properties of pull--backed ones 
of universal bundles over the infinite--dimensional Grassmann manifolds 
and Chern characters, see \cite{MN}.

Let $\calh$ be a separable Hilbert space over $\fukuso$.
For $m\in{\bf N}$, we set
\begin{equation}
  \label{eq:stmh}
  \kansu{\stm}{\cal H}
  \equiv
  \left\{
    V=\left(v_1,\cdots,v_m\right)
    \in
    \calh\times\cdots\times\calh\ \vert\  V^\dagger V \in GL(m;\fukuso)
  \right\}\ .
\end{equation}
This is called a (universal) Stiefel manifold.
Note that the unitary group $U(m)$ acts on $\kansu{\stm}{\calh}$
from the right :
\begin{equation}
  \label{eq:stmsha}
  \kansu{\stm}{\calh}\times\kansu{U}{m}
  \longrightarrow
  \kansu{\stm}{\calh}\  :\  \left( V,a\right)\longmapsto Va\ .
\end{equation}
Next we define a (universal) Grassmann manifold
\begin{equation}
  \kansu{\grm}{\calh}
  \equiv
  \left\{
    X\in\kansu{M}{\calh}\ \vert\ 
    X^2=X, X^\dagger=X\  \mathrm{and}\  \mathrm{tr}X=m\right\}\ ,
\end{equation}
where $M(\calh)$ denotes a space of all bounded linear operators on $\calh$.
Then we have a projection
\begin{equation}
  \label{eq:piteigi}
  \pi : \kansu{\stm}{\calh}\longrightarrow\kansu{\grm}{\calh}\ ,
  \quad \kansu{\pi}{V}\equiv V(V^{\dagger}V)^{-1}V^\dagger\ ,
\end{equation}
compatible with the action (\ref{eq:stmsha}) 
($\kansu{\pi}{Va}=Va\{a^{-1}(V^{\dagger}V)^{-1}a\}(Va)^\dagger
=\kansu{\pi}{V}$).

Now the set
\begin{equation}
  \label{eq:principal}
  \left\{
    \kansu{U}{m}, \kansu{\stm}{\calh}, \pi, \kansu{\grm}{\calh}
  \right\}\ ,
\end{equation}
is called a (universal) principal $U(m)$ bundle, 
see \cite{MN} and \cite{KF1}. \quad We set
\begin{equation}
  \label{eq:emh}
  \kansu{\eem}{\cal H}
  \equiv
  \left\{
    \left(X,v\right)
    \in
    \kansu{\grm}{\calh}\times\calh\ \vert\  Xv=v \right\}\ .
\end{equation}
Then we have also a projection 
\begin{equation}
  \label{eq:piemgrm}
  \pi : \kansu{\eem}{\calh}\longrightarrow\kansu{\grm}{\calh}\ ,
  \quad \kansu{\pi}{\left(X,v\right)}\equiv X\ .
\end{equation}
The set
\begin{equation}
  \label{eq:universal}
  \left\{
    \fukuso^m, \kansu{\eem}{\calh}, \pi, \kansu{\grm}{\calh}
  \right\}\ ,
\end{equation}
is called a (universal) $m$--th vector bundle. This vector bundle is 
one associated with the principal $U(m)$ bundle (\ref{eq:principal}). 

Next let ${\calm}$ be a finite or infinite dimensional differentiable manifold 
and the map 
\begin{equation}
  \label{eq:projector}
   P : {\calm} \longrightarrow \kansu{\grm}{\calh}
\end{equation} 
be given (called a projector). Using this $P$ we can make the bundles 
(\ref{eq:principal}) and (\ref{eq:universal}) pullback over ${\calm}$ :
\begin{eqnarray}
  \label{eq:hikimodoshi1}
  &&\left\{\kansu{U}{m},\widetilde{St}, \pi_{\widetilde{St}}, {\calm}\right\}
  \equiv
  P^*\left\{\kansu{U}{m}, \kansu{\stm}{\calh}, \pi, 
  \kansu{\grm}{\calh}\right\}
  \ , \\
  \label{eq:hikimodoshi2}
  &&\left\{\fukuso^m,\widetilde{E}, \pi_{\widetilde{E}}, {\calm}\right\}
  \equiv
  P^*\left\{\fukuso^m, \kansu{\eem}{\calh}, \pi, \kansu{\grm}{\calh}\right\}
  \ , 
\end{eqnarray}

\[    
   \matrix{
    \kansu{U}{m}&&\kansu{U}{m}\cr
    \Big\downarrow&&\Big\downarrow\cr
    \widetilde{St}&\longrightarrow&\kansu{\stm}{\calh}\cr
    \Big\downarrow&&\Big\downarrow\cr
    {\calm}&\stackrel{P}{\longrightarrow}&\kansu{\grm}{\calh}\cr
           } \qquad \qquad  
   \matrix{
    \fukuso^m&&\fukuso^m\cr
    \Big\downarrow&&\Big\downarrow\cr
    \widetilde{E}&\longrightarrow&\kansu{E_m}{\calh}\cr
    \Big\downarrow&&\Big\downarrow\cr
    {\calm}&\stackrel{P}{\longrightarrow}&\kansu{\grm}{\calh}\cr
           } 
\]
\par \vspace{5mm} \noindent
see \cite{MN}. (\ref{eq:hikimodoshi2}) is of course a vector bundle 
associated with (\ref{eq:hikimodoshi1}).

For this bundle the (global) curvature ($2$--) form $\bf\Omega$ is given by 
\begin{equation}
  \label{eq:curvature}
  {\bf\Omega}=PdP\wedge dP 
\end{equation}
making use of (\ref{eq:projector}), where $d$ is the usual differential form 
on $\bf\Omega$. 
For the bundles Chern characters play an essential role in several geometric 
properties. In this case Chern characters are given by 
\begin{equation}
  \label{eq:Chern--classes}
  {\bf\Omega},\ {\bf\Omega}^2,\ \cdots,\ {\bf\Omega}^{m/2}; \quad 
  {\bf\Omega}^2={\bf\Omega}\wedge{\bf\Omega},\ \mbox{etc}, 
\end{equation}
where we have assumed that $m=\mbox{dim}{\calm}$ is even. 
In this paper we don't take the trace of (\ref{eq:Chern--classes}), so 
it may be better to call them densities for Chern characters.  

To calculate these quantities in infinite--dimensional cases is not so 
easy. In the next section let us calculate these ones in the special cases.

We now define our projectors for the latter aim. In the following $
\calh=\calh_{K}$. For $w_1, w_2, \cdots, 
w_m \in \fukuso$ we consider a set of generalized coherent states 
$\{\ket{w_1}, \ket{w_2}, \cdots,\ket{w_m}\}$ and set 
\begin{equation}
  \label{eq:a set of coherent states}   
   V_{m}({\bf w})=(\ket{w_1},\ket{w_2}, \cdots, \ket{w_m})\equiv V_{m}
\end{equation}
where ${\bf w}=(w_1, w_2, \cdots, w_m)$. 
Since ${V_{m}}^{\dagger}V_{m}=(\braket{w_i}{w_j}) \in M(m,\fukuso)$,  
we define 
\begin{equation}
  \label{eq:domain}
    {\cal D}_m \equiv \{{\bf w}\in \fukuso^{m}\ \vert\ 
               \mbox{det}({V_{m}}^{\dagger}V_{m}) \ne 0 \}. 
\end{equation}
For example ${V_{1}}^{\dagger}V_{1}=1$ for $m=1$, and for $m=2$ 
\[
   \mbox{det}({V_{2}}^{\dagger}V_{2}) = 
  \left|
    \begin{array}{cc}
        1 & a \\
        \bar{a} & 1 
    \end{array}
  \right|
   = 1-\vert{a}\vert^{2} \geq 0
\]
where $a=\braket{w_1}{w_2}$. So from (\ref{eq:futou-shiki-3}) we have 
\begin{eqnarray}
  \label{eq:condition-1}   
   {\cal D}_1 &=&\{w \in \fukuso \ \vert\ \mbox{no conditions}\} 
               = \fukuso ,  \\
  \label{eq:condition-2}   
   {\cal D}_2 &=&\{(w_1,w_2) \in \fukuso^2 \ \vert\ w_1\ne w_2 \}.
\end{eqnarray}
For ${\cal D}_m\ (m\geq 3)$ it is not easy for us to give a simple 
condition like (\ref{eq:condition-2}). At any rate 
$
V_{m} \in \kansu{\stm}{\calh} \ \mbox{for}\ {\bf w} \in {\cal D}_m 
$ .
Now let us define our projector $P$ as follows : 
\begin{eqnarray}
  \label{eq:real-projector}
  P : {\cal D}_m \longrightarrow \kansu{\grm}{\calh}\ , \quad 
    P({\bf w})= V_{m}(V_{m}^{\dagger}V_{m})^{-1}V_{m}^{\dagger}\ .
\end{eqnarray}
In the following we set $V=V_{m}$ for simplicity.  Let us calculate 
(\ref{eq:curvature}). Since 
\begin{equation}
    dP=V(V^{\dagger}V)^{-1}dV^{\dagger}
     \{
        {\bf 1}-V(V^{\dagger}V)^{-1}V^{\dagger}
     \}
     + \{{\bf 1}-V(V^{\dagger}V)^{-1}V^{\dagger}\}dV(V^{\dagger}V)^{-1}
      V^{\dagger} 
\end{equation}
where $d=\sum_{j=1}^{m}\left(dw_{j}\frac{\partial}{\partial w_{j}}+ 
d{\bar w_{j}}\frac{\partial}{\partial {\bar w_{j}}}\right)$, we have 
\[
  PdP=V(V^{\dagger}V)^{-1}dV^{\dagger}
      \{{\bf 1}-V(V^{\dagger}V)^{-1}V^{\dagger}\} 
\]
after some calculation. Therefore we obtain  
\begin{equation}
  \label{eq:curvature-local}
  PdP\wedge dP=V(V^{\dagger}V)^{-1}[dV^{\dagger}\{
               {\bf 1}-V(V^{\dagger}V)^{-1}V^{\dagger}
               \}dV](V^{\dagger}V)^{-1}V^{\dagger}\ .
\end{equation}
Our main calculation is $dV^{\dagger}
\{{\bf 1}-V(V^{\dagger}V)^{-1}V^{\dagger}\}dV$, which is rewritten as 
\begin{equation}
  \label{eq:curvature-decomposition}
  dV^{\dagger}\{{\bf 1}-V(V^{\dagger}V)^{-1}V^{\dagger}\}dV=
  {[\{{\bf 1}-V(V^{\dagger}V)^{-1}V^{\dagger}\}dV]}^{\dagger}\ 
  [\{{\bf 1}-V(V^{\dagger}V)^{-1}V^{\dagger}\}dV] 
\end{equation}
since $Q \equiv {\bf 1}-V(V^{\dagger}V)^{-1}V^{\dagger}$ is also a projector
($Q^2=Q$ and $Q^{\dagger}=Q$). Therefore the first step for us is to 
calculate the term 
\begin{equation}
  \label{eq:main-term}
  \{{\bf 1}-V(V^{\dagger}V)^{-1}V^{\dagger}\}dV\ .
\end{equation}
\par \noindent 
Let us summarize {\bf our process of calculations} : 
\begin{eqnarray}
  &&\mbox{1--st step}\qquad  \{{\bf 1}-V(V^{\dagger}V)^{-1}V^{\dagger}\}dV\ 
     \cdots (\ref {eq:main-term}),  \nonumber \\
  &&\mbox{2--nd step}\qquad 
    dV^{\dagger}\{{\bf 1}-V(V^{\dagger}V)^{-1}V^{\dagger}\}dV\  \cdots 
    (\ref{eq:curvature-decomposition}),  \nonumber \\
  &&\mbox{3--rd step}\qquad 
    V(V^{\dagger}V)^{-1}
    [dV^{\dagger}\{{\bf 1}-V(V^{\dagger}V)^{-1}V^{\dagger}\}dV]   
    (V^{\dagger}V)^{-1}V^{\dagger}\  \cdots 
    (\ref{eq:curvature-local}).  \nonumber 
\end{eqnarray}

\vspace{2cm}
\section{Calculations of Chern Characters}

In this section we calculate the Chern characters only for the cases $m=1$
and $m=2$. Even for $m=2$ the calculation is complicated enough. 
For $m\geq 3$ calculations may become miserable.

\subsection{M=1}

In this case $\braket{w}{w}=1$, so our projector is very simple to be 
\begin{equation}
  \label{eq:1-projector} 
     P(w)=\ket{w}\bra{w}. 
\end{equation}
In this case the calculation of curvature is relatively simple. 
From (\ref{eq:curvature-local}) we have 
\begin{equation}
  \label{eq:1-curvature} 
  PdP\wedge dP=\ket{w}\{d\bra{w}({\bf 1}_{K}-\ket{w}\bra{w})d\ket{w}\}\bra{w}
         =\ket{w}\bra{w}\{d\bra{w}({\bf 1}_{K}-\ket{w}\bra{w})d\ket{w}\}, 
\end{equation}
where $d=dw\frac{\partial}{\partial w}+d{\bar w}\frac{\partial}
{\partial {\bar w}}$. 
Since $\ket{w}=(1-\zettai{\zeta}^2)^{K}\mbox{exp}(\zeta K_{+})\kett{K}{0}$
by (\ref{eq:2-2-right}), 
\begin{equation}
  \label{eq:bibun-koushiki} 
 d\ket{w}=\left\{d\zeta K_{+}+Kd\mbox{log}(1-\zettai{\zeta}^2)\right\}\ket{w}
\end{equation}
by some calculation, so that 
\begin{eqnarray}
  ({\bf 1}_{K}-\ket{w}\bra{w})d\ket{w}&=&
    ({\bf 1}_{K}-\ket{w}\bra{w})K_{+}\ket{w}d\zeta =
    (K_{+}-\bra{w}K_{+}\ket{w})\ket{w}d\zeta  \nonumber \\
  &=&\left(K_{+}-\frac{2K{\bar \zeta}}{1-\zettai{\zeta}^2}\right)d\zeta\ket{w}
\end{eqnarray}
because $({\bf 1}_{K}-\ket{w}\bra{w})\ket{w}={\bf 0}$. Similarly we have 
\begin{equation}
 d\bra{w}({\bf 1}_{K}-\ket{w}\bra{w})=\bra{w}
 \left(K_{-}-\frac{2K{\zeta}}{1-\zettai{\zeta}^2}\right)d{\bar \zeta}
\end{equation} 
Now we are in a position to determine the curvature form 
(\ref{eq:1-curvature}).
\begin{eqnarray}
  &&d\bra{w}({\bf 1}_{K}-\ket{w}\bra{w})d\ket{w}    \nonumber \\
  &&=\bra{w}
     \left(K_{-}-\frac{2K{\zeta}}{1-\zettai{\zeta}^2}\right)
     \left(K_{+}-\frac{2K{\bar \zeta}}{1-\zettai{\zeta}^2}\right)
     \ket{w} d{\bar \zeta}\wedge d\zeta   \nonumber \\
 &&=\left\{\bra{w}K_{-}K_{+}\ket{w}
           -\frac{2K{\bar \zeta}}{1-\zettai{\zeta}^2}\bra{w}K_{-}\ket{w}
           -\frac{2K{\zeta}}{1-\zettai{\zeta}^2}\bra{w}K_{+}\ket{w}
           +\frac{4K^{2}\zettai{\zeta}^2}{(1-\zettai{\zeta}^2)^2}
    \right\}d{\bar \zeta}\wedge d\zeta  \nonumber \\
 &&=\frac{2K}{(1-\zettai{\zeta}^2)^2}d{\bar \zeta}\wedge d\zeta
\end{eqnarray}
after some algebra with (\ref{eq:kitaichi-1}) and (\ref{eq:kitaichi-2}). 
Therefore
\begin{equation}
  \label{eq:result-1}
 {\bf\Omega}= PdP\wedge dP=\ket{w}\bra{w}
              \frac{2Kd{\bar \zeta}\wedge d\zeta}
                   {(1-\zettai{\zeta}^2)^2}.
\end{equation}
From this result we know 
\[
  \frac{{\bf\Omega}}{2\pi i}=\frac{2K}{\pi}
    \frac{d{\zeta_1}\wedge d{\zeta_2}}{(1-\zettai{\zeta}^2)^2}\ket{w}\bra{w}
  =\frac{2K}{\pi}\frac{d{\zeta_1}\wedge d{\zeta_2}}{(1-\zettai{\zeta}^2)^2}
   \ket{\zeta}\bra{\zeta}
\]
by (\ref{eq:2-2-right}) 
when $\zeta=\zeta_{1}+\sqrt{-1}\zeta_{2}$. If we define a constant  
\begin{equation}
     C_{K}=\frac{2K-1}{2K},
\end{equation}
then we have 
\begin{equation}
     C_{K}\frac{{\bf\Omega}}{2\pi i}=\frac{2K-1}{\pi}
  \frac{d{\zeta_1}\wedge d{\zeta_2}}{(1-\zettai{\zeta}^2)^2}
  \ket{\zeta}\bra{\zeta}.  
\end{equation}
This gives the resolution of unity in (\ref{eq:2-2-18}). But the situation 
is a bit different from \cite{KF9} in which the constant corresponding to 
$C_{K}$ is just one.  
\begin{flushleft}
{\bf Problem}\quad What is a (deep) meaning of $C_{K}$  ? 
\end{flushleft}

\vspace{10mm}
\subsection{M=2 $\cdots$ Main Result}

First of all let us determine the projector. Since $V=(\ket{w_1},\ket{w_2})$ 
we have easily 
\begin{eqnarray}
   P(w_1, w_2)&=&\left(\ket{w_1},\ket{w_2}\right)
 {\left(
    \begin{array}{cc}
        1 & \braket{w_1}{w_2} \\
        \braket{w_2}{w_1} & 1 
    \end{array}
  \right)}^{-1}
 {\left(
    \begin{array}{c}
       \bra{w_1} \\
       \bra{w_2}  
    \end{array}
  \right)}   \nonumber \\
 &=&\frac{1}{1-\vert{\braket{w_1}{w_2}}\vert^{2}}
   (\ket{w_1},\ket{w_2})
 {\left(
    \begin{array}{cc}
        1 & -\braket{w_1}{w_2} \\
        -\braket{w_2}{w_1} & 1 
    \end{array}
  \right)}
 {\left(
    \begin{array}{c}
       \bra{w_1} \\
       \bra{w_2}  
    \end{array}
  \right)}   \nonumber \\
 &=&\frac{1}{1-\vert{\braket{w_1}{w_2}}\vert^{2}}
   \left(\ket{w_1}\bra{w_1}-\braket{w_2}{w_1}\ket{w_2}\bra{w_1}
         -\braket{w_1}{w_2}\ket{w_1}\bra{w_2}+\ket{w_2}\bra{w_2}
   \right). \nonumber \\
 &&{}
\end{eqnarray}
\par \noindent
Let us calculate (\ref{eq:main-term}) : Since 
\begin{eqnarray}
  dV&=&(d\ket{w_1},d\ket{w_2}) \nonumber \\
&=&
\left(
 \left\{d\zeta_{1}K_{+}+Kd\mbox{log}(1-\zettai{\zeta_{1}}^2)\right\}\ket{w_1}, 
 \left\{d\zeta_{2}K_{+}+Kd\mbox{log}(1-\zettai{\zeta_{2}}^2)\right\}\ket{w_2} 
\right)     \nonumber
\end{eqnarray}
by (\ref{eq:bibun-koushiki}) where $d=\sum_{j=1}^{2}\left(
dw_j\frac{\partial}{\partial w_j}+d{\bar w_j}\frac{\partial}
{\partial {\bar w_j}} \right)$, 
the long but straightforward calculation leads to  
\begin{equation}
  \{{\bf 1}_{K}-V(V^{\dagger}V)^{-1}V^{\dagger}\}dV=(G_1, G_2) 
\end{equation}
where 
\begin{eqnarray}
  G_1&=&\left\{
         \left(K_{+}-\frac{2K{\bar \zeta_2}}{1-{\bar \zeta_2}\zeta_1}
         \right)
         - \frac{2K}{1-\zettai{\braket{w_1}{w_2}}^2}
           \frac{{\bar \zeta_1}-{\bar \zeta_2}}
                {(1-\zettai{\zeta_1}^2)(1-{\bar \zeta_2}\zeta_1)}
        \right\}d\zeta_1 \ket{w_1}      \nonumber \\
     &&+ \frac{2K\braket{w_2}{w_1}}{1-\zettai{\braket{w_1}{w_2}}^2}
           \frac{{\bar \zeta_1}-{\bar \zeta_2}}
                {(1-\zettai{\zeta_1}^2)(1-{\bar \zeta_2}\zeta_1)}
                d\zeta_1 \ket{w_2}\ ,    \nonumber  \\
&&{}  \nonumber \\
  G_2&=& \frac{2K\braket{w_1}{w_2}}{1-\zettai{\braket{w_1}{w_2}}^2}
           \frac{{\bar \zeta_2}-{\bar \zeta_1}}
                {(1-{\bar \zeta_1}\zeta_2)(1-\zettai{\zeta_2}^2)}
                d\zeta_2 \ket{w_1}      \nonumber \\
      &&+ \left\{
           \left(K_{+}-\frac{2K{\bar \zeta_1}}{1-{\bar \zeta_1}\zeta_2}
           \right)
         - \frac{2K}{1-\zettai{\braket{w_1}{w_2}}^2}
           \frac{{\bar \zeta_2}-{\bar \zeta_1}}
                {(1-\zettai{\zeta_2}^2)(1-{\bar \zeta_1}\zeta_2)}
        \right\}d\zeta_2 \ket{w_2}\ .  
\end{eqnarray}
Therefore by (\ref{eq:curvature-decomposition})
\begin{equation}
  \label{eq:matrix-form}
  dV^{\dagger}\{{\bf 1}_{K}-V(V^{\dagger}V)^{-1}V^{\dagger}\}dV=
  \left(
    \begin{array}{cc}
       F_{11}& F_{12} \\
       F_{21}& F_{22} 
    \end{array}
  \right)  
\end{equation}
where 
\begin{eqnarray}
  F_{11}&=&\frac{2K}{(1-\zettai{\zeta_1}^2)^2}
    \left\{1- \frac{\zettai{\braket{w_1}{w_2}}^2}
                   {1-\zettai{\braket{w_1}{w_2}}^2}
              \frac{2K\zettai{\zeta_1-\zeta_2}^2}
                   {\zettai{1-{\bar \zeta_1}\zeta_2}^2}
    \right\}d{\bar \zeta_1}\wedge d\zeta_1\ ,  \nonumber \\
&&{} \nonumber \\
  F_{12}&=&\frac{2K\braket{w_1}{w_2}}{(1-{\bar \zeta_1}\zeta_2)^2}
    \left\{1- \frac{1}
                   {1-\zettai{\braket{w_1}{w_2}}^2}
              \frac{2K\zettai{\zeta_1-\zeta_2}^2}
                   {(1-\zettai{\zeta_1}^2)(1-\zettai{\zeta_2}^2)}
    \right\}d{\bar \zeta_1}\wedge d\zeta_2\ ,  \nonumber \\
&&{} \nonumber \\
  F_{21}&=&\frac{2K\braket{w_2}{w_1}}{(1-{\bar \zeta_2}\zeta_1)^2}
    \left\{1- \frac{1}
                   {1-\zettai{\braket{w_1}{w_2}}^2}
              \frac{2K\zettai{\zeta_1-\zeta_2}^2}
                   {(1-\zettai{\zeta_1}^2)(1-\zettai{\zeta_2}^2)}
    \right\}d{\bar \zeta_2}\wedge d\zeta_1\ ,  \nonumber \\
&&{} \nonumber \\
  F_{22}&=&\frac{2K}{(1-\zettai{\zeta_2}^2)^2}
    \left\{1- \frac{\zettai{\braket{w_1}{w_2}}^2}
                   {1-\zettai{\braket{w_1}{w_2}}^2}
              \frac{2K\zettai{\zeta_1-\zeta_2}^2}
                   {\zettai{1-{\bar \zeta_2}\zeta_1}^2}
    \right\}d{\bar \zeta_2}\wedge d\zeta_2\ .  \\
&&{} \nonumber 
\end{eqnarray}
\par \noindent \vspace{5mm}
Now we are in a position to determine the curvature form 
(\ref{eq:curvature-local}). 
Since 
\begin{eqnarray}
  V(V^{\dagger}V)^{-1}&=&
   \frac{1}{1-\vert{\braket{w_1}{w_2}}\vert^{2}}
   (\ket{w_1}-\braket{w_2}{w_1}\ket{w_2}, 
      \ket{w_2}-\braket{w_1}{w_2}\ket{w_1}), \nonumber \\
 (V^{\dagger}V)^{-1}V^{\dagger}&=&
   \frac{1}{1-\vert{\braket{w_1}{w_2}}\vert^{2}}
  {\left(
    \begin{array}{c}
       \bra{w_1}- \bra{w_2}\braket{w_1}{w_2}\\
       \bra{w_2}- \bra{w_1}\braket{w_2}{w_1} 
    \end{array}
  \right)}  \nonumber   
\end{eqnarray}
we obtain after some algebra 
\begin{eqnarray}
  \label{eq:exact-calculation-2form} 
 &{\bf\Omega}&=PdP\wedge dP   \nonumber \\
 &=&\frac{1}{(1-\vert{\braket{w_1}{w_2}}\vert^{2})^{2}}
   \left(
           \ket{w_1}\bra{w_1}L_1
          -\braket{w_2}{w_1}\ket{w_2}\bra{w_1}L_2
          -\braket{w_1}{w_2}\ket{w_1}\bra{w_2}L_3
          +\ket{w_2}\bra{w_2}L_4
   \right),  \nonumber \\
&&{} 
\end{eqnarray}
where 
\begin{eqnarray}
  L_1 &=&\frac{2K}{(1-\zettai{\zeta_1}^2)^2}
      \left\{1-\frac{\zettai{\braket{w_1}{w_2}}^2}
                    {1-\zettai{\braket{w_1}{w_2}}^2}
               \frac{2K\zettai{\zeta_1-\zeta_2}^2}
                    {\zettai{1-{\bar \zeta_1}\zeta_2}^2}
      \right\}d{\bar \zeta_1}\wedge d\zeta_1       \nonumber \\
      &-&\frac{2K\zettai{\braket{w_1}{w_2}}^2}{(1-{\bar \zeta_2}\zeta_1)^2}
      \left\{1-\frac{1}
                    {1-\zettai{\braket{w_1}{w_2}}^2}
               \frac{2K\zettai{\zeta_1-\zeta_2}^2}
                    {(1-\zettai{\zeta_1}^2)(1-\zettai{\zeta_2}^2)}
      \right\}d{\bar \zeta_2}\wedge d\zeta_1       \nonumber \\
      &-&\frac{2K\zettai{\braket{w_1}{w_2}}^2}{(1-{\bar \zeta_1}\zeta_2)^2}
      \left\{1-\frac{1}
                    {1-\zettai{\braket{w_1}{w_2}}^2}
               \frac{2K\zettai{\zeta_1-\zeta_2}^2}
                    {(1-\zettai{\zeta_1}^2)(1-\zettai{\zeta_2}^2)}
      \right\}d{\bar \zeta_1}\wedge d\zeta_2       \nonumber \\
      &+&\frac{2K\zettai{\braket{w_1}{w_2}}^2}{(1-\zettai{\zeta_2}^2)^2}
      \left\{1-\frac{\zettai{\braket{w_1}{w_2}}^2}
                    {1-\zettai{\braket{w_1}{w_2}}^2}
               \frac{2K\zettai{\zeta_1-\zeta_2}^2}
                    {\zettai{1-{\bar \zeta_1}\zeta_2}^2}
      \right\}d{\bar \zeta_2}\wedge d\zeta_2\ ,     \nonumber \\
&&{} \nonumber \\
  L_2 &=&\frac{2K}{(1-\zettai{\zeta_1}^2)^2}
      \left\{1-\frac{\zettai{\braket{w_1}{w_2}}^2}
                    {1-\zettai{\braket{w_1}{w_2}}^2}
               \frac{2K\zettai{\zeta_1-\zeta_2}^2}
                    {\zettai{1-{\bar \zeta_1}\zeta_2}^2}
      \right\}d{\bar \zeta_1}\wedge d\zeta_1       \nonumber \\
      &-&\frac{2K}{(1-{\bar \zeta_2}\zeta_1)^2}
      \left\{1-\frac{1}
                    {1-\zettai{\braket{w_1}{w_2}}^2}
               \frac{2K\zettai{\zeta_1-\zeta_2}^2}
                    {(1-\zettai{\zeta_1}^2)(1-\zettai{\zeta_2}^2)}
      \right\}d{\bar \zeta_2}\wedge d\zeta_1       \nonumber \\
      &-&\frac{2K\zettai{\braket{w_1}{w_2}}^2}{(1-{\bar \zeta_1}\zeta_2)^2}
      \left\{1-\frac{1}
                    {1-\zettai{\braket{w_1}{w_2}}^2}
               \frac{2K\zettai{\zeta_1-\zeta_2}^2}
                    {(1-\zettai{\zeta_1}^2)(1-\zettai{\zeta_2}^2)}
      \right\}d{\bar \zeta_1}\wedge d\zeta_2       \nonumber \\
      &+&\frac{2K}{(1-\zettai{\zeta_2}^2)^2}
      \left\{1-\frac{\zettai{\braket{w_1}{w_2}}^2}
                    {1-\zettai{\braket{w_1}{w_2}}^2}
               \frac{2K\zettai{\zeta_1-\zeta_2}^2}
                    {\zettai{1-{\bar \zeta_1}\zeta_2}^2}
      \right\}d{\bar \zeta_2}\wedge d\zeta_2\ ,     \nonumber  \\
&&{} \nonumber \\
  L_3 &=&\frac{2K}{(1-\zettai{\zeta_1}^2)^2}
      \left\{1-\frac{\zettai{\braket{w_1}{w_2}}^2}
                    {1-\zettai{\braket{w_1}{w_2}}^2}
               \frac{2K\zettai{\zeta_1-\zeta_2}^2}
                    {\zettai{1-{\bar \zeta_1}\zeta_2}^2}
      \right\}d{\bar \zeta_1}\wedge d\zeta_1       \nonumber \\
      &-&\frac{2K\zettai{\braket{w_1}{w_2}}^2}{(1-{\bar \zeta_2}\zeta_1)^2}
      \left\{1-\frac{1}
                    {1-\zettai{\braket{w_1}{w_2}}^2}
               \frac{2K\zettai{\zeta_1-\zeta_2}^2}
                    {(1-\zettai{\zeta_1}^2)(1-\zettai{\zeta_2}^2)}
      \right\}d{\bar \zeta_2}\wedge d\zeta_1       \nonumber \\
      &-&\frac{2K}{(1-{\bar \zeta_1}\zeta_2)^2}
      \left\{1-\frac{1}
                    {1-\zettai{\braket{w_1}{w_2}}^2}
               \frac{2K\zettai{\zeta_1-\zeta_2}^2}
                    {(1-\zettai{\zeta_1}^2)(1-\zettai{\zeta_2}^2)}
      \right\}d{\bar \zeta_1}\wedge d\zeta_2       \nonumber \\
      &+&\frac{2K}{(1-\zettai{\zeta_2}^2)^2}
      \left\{1-\frac{\zettai{\braket{w_1}{w_2}}^2}
                    {1-\zettai{\braket{w_1}{w_2}}^2}
               \frac{2K\zettai{\zeta_1-\zeta_2}^2}
                    {\zettai{1-{\bar \zeta_1}\zeta_2}^2}
      \right\}d{\bar \zeta_2}\wedge d\zeta_2\ ,     \nonumber  \\
&&{} \nonumber \\
  L_4 &=&\frac{2K\zettai{\braket{w_1}{w_2}}^2}{(1-\zettai{\zeta_1}^2)^2}
      \left\{1-\frac{\zettai{\braket{w_1}{w_2}}^2}
                    {1-\zettai{\braket{w_1}{w_2}}^2}
               \frac{2K\zettai{\zeta_1-\zeta_2}^2}
                    {\zettai{1-{\bar \zeta_1}\zeta_2}^2}
      \right\}d{\bar \zeta_1}\wedge d\zeta_1       \nonumber \\
      &-&\frac{2K\zettai{\braket{w_1}{w_2}}^2}{(1-{\bar \zeta_2}\zeta_1)^2}
      \left\{1-\frac{1}
                    {1-\zettai{\braket{w_1}{w_2}}^2}
               \frac{2K\zettai{\zeta_1-\zeta_2}^2}
                    {(1-\zettai{\zeta_1}^2)(1-\zettai{\zeta_2}^2)}
      \right\}d{\bar \zeta_2}\wedge d\zeta_1       \nonumber \\
      &-&\frac{2K\zettai{\braket{w_1}{w_2}}^2}{(1-{\bar \zeta_1}\zeta_2)^2}
      \left\{1-\frac{1}
                    {1-\zettai{\braket{w_1}{w_2}}^2}
               \frac{2K\zettai{\zeta_1-\zeta_2}^2}
                    {(1-\zettai{\zeta_1}^2)(1-\zettai{\zeta_2}^2)}
      \right\}d{\bar \zeta_1}\wedge d\zeta_2       \nonumber \\
      &+&\frac{2K}{(1-\zettai{\zeta_2}^2)^2}
      \left\{1-\frac{\zettai{\braket{w_1}{w_2}}^2}
                    {1-\zettai{\braket{w_1}{w_2}}^2}
               \frac{2K\zettai{\zeta_1-\zeta_2}^2}
                    {\zettai{1-{\bar \zeta_1}\zeta_2}^2}
      \right\}d{\bar \zeta_2}\wedge d\zeta_2\ .           \\
&&{} \nonumber 
\end{eqnarray}
\par \vspace{5mm} \noindent 
This is our main result. Next let us calculate ${\bf\Omega}^2$ (${\bf\Omega}^k
={\bf 0}$ for $k\geq 3$) : From (\ref{eq:exact-calculation-2form}) we obtain 
after long calculation 
\vspace{5mm} 
\begin{eqnarray}
 \label{eq:exact-calculation-4form}
 {\bf\Omega}^2 &=&
 \frac{1}{(1-\vert{\braket{w_1}{w_2}}\vert^{2})^{4}}  
 \left(
           \ket{w_1}\bra{w_1}M_1
          -\braket{w_2}{w_1}\ket{w_2}\bra{w_1}M_2 
          -\braket{w_1}{w_2}\ket{w_1}\bra{w_2}M_3
          +\ket{w_2}\bra{w_2}M_4
 \right) \nonumber \\
 &\times&  
   d{\bar \zeta_1}\wedge d\zeta_1\wedge d{\bar \zeta_2}\wedge d\zeta_2 \ ,  
\end{eqnarray}
where 
\begin{eqnarray}
 M_1&=&\frac{4K^2\zettai{\braket{w_1}{w_2}}^2}
            {(1-\zettai{\zeta_1}^2)^2(1-\zettai{\zeta_2}^2)^2
             \zettai{1-{\bar \zeta_1}\zeta_2}^4}\ [  \nonumber \\
   &&\ \left\{2\zettai{1-{\bar \zeta_1}\zeta_2}^4
             (1-\zettai{\braket{w_1}{w_2}}^2)
           -(1-\zettai{\zeta_1}^2)^2(1-\zettai{\zeta_2}^2)^2
            (1-\zettai{\braket{w_1}{w_2}}^4)
      \right\} \nonumber \\
  &&\ -2 \left\{2\zettai{1-{\bar \zeta_1}\zeta_2}^{2}
              \zettai{\braket{w_1}{w_2}}^{2} 
          -(1-\zettai{\zeta_1}^2)(1-\zettai{\zeta_2}^2)
           (1+\zettai{\braket{w_1}{w_2}}^2)
        \right\} 2K\zettai{\zeta_1-\zeta_2}^2   \nonumber \\
  &&\ -(1+2\zettai{\braket{w_1}{w_2}}^2)4K^{2}\zettai{\zeta_1-\zeta_2}^{4}\ 
  ], \nonumber \\
&&{} \nonumber \\
 M_2&=&\frac{4K^2}
            {(1-\zettai{\zeta_1}^2)^2(1-\zettai{\zeta_2}^2)^2
             \zettai{1-{\bar \zeta_1}\zeta_2}^4}\ [  \nonumber \\
   &&\ \left\{\zettai{1-{\bar \zeta_1}\zeta_2}^4
             (1-\zettai{\braket{w_1}{w_2}}^4)
           -2(1-\zettai{\zeta_1}^2)^2(1-\zettai{\zeta_2}^2)^2
             \zettai{\braket{w_1}{w_2}}^2(1-\zettai{\braket{w_1}{w_2}}^2)
      \right\} \nonumber \\
  &&\ -2 \left\{\zettai{1-{\bar \zeta_1}\zeta_2}^{2}
           \zettai{\braket{w_1}{w_2}}^2(1+\zettai{\braket{w_1}{w_2}}^{2}) 
          -2(1-\zettai{\zeta_1}^2)(1-\zettai{\zeta_2}^2)
            \zettai{\braket{w_1}{w_2}}^2
        \right\} 2K\zettai{\zeta_1-\zeta_2}^2   \nonumber \\
  &&\ -\zettai{\braket{w_1}{w_2}}^{2}(2+\zettai{\braket{w_1}{w_2}}^2)
       4K^{2}\zettai{\zeta_1-\zeta_2}^{4}\ 
  ], \nonumber \\
&&{}  \nonumber \\
M_3&=&M_2, \quad \mbox{and}\quad  M_4=M_1\ .
\end{eqnarray}
This is a second main result in this paper.

\par \noindent
We have calculated the Chern characters for $m=2$. Since our results are 
in a certain sense ``raw'' (remember that we have not taken the trace), one 
can freely ``cook'' them. We leave it to the readers.

\vspace{1cm}
\subsection{Problems}
 
Before concluding this section let us propose problems.  
\begin{flushleft}
{\bf Problem\ 3}\quad For the case of $m=3$ perform the similar calculations ! 
\end{flushleft}
We want to calculate them up to this case. Therefore let us give an explicit 
form to the projector : 
\begin{eqnarray}
   &&P(w_1, w_2, w_3)=\left(\ket{w_1},\ket{w_2},\ket{w_3}\right)
 {\left(
    \begin{array}{ccc}
        1 & \braket{w_1}{w_2} & \braket{w_1}{w_3} \\
        \braket{w_2}{w_1} & 1 & \braket{w_2}{w_3} \\
        \braket{w_3}{w_1} & \braket{w_3}{w_2} & 1 
    \end{array}
  \right)}^{-1}
 {\left(
    \begin{array}{c}
       \bra{w_1} \\
       \bra{w_2} \\
       \bra{w_3}  
    \end{array}
  \right)}   \nonumber \\
    &&\quad =\frac{1}{\mbox{det}M}
     {\bf [}\ 
      \{1-\szetta{\braket{w_2}{w_3}}\}\ket{w_1}\bra{w_1}
     -\{\braket{w_1}{w_2}-\braket{w_1}{w_3}\braket{w_3}{w_2}\}
        \ket{w_1}\bra{w_2}  \nonumber \\
    &&\quad\ -\{\braket{w_1}{w_3}-\braket{w_1}{w_2}\braket{w_2}{w_3}\}
        \ket{w_1}\bra{w_3}
      -\{\braket{w_2}{w_1}-\braket{w_2}{w_3}\braket{w_3}{w_1}\}
        \ket{w_2}\bra{w_1}  \nonumber \\
    &&\quad\  + \{1-\szetta{\braket{w_1}{w_3}}\}\ket{w_2}\bra{w_2}
       -\{\braket{w_2}{w_3}-\braket{w_2}{w_1}\braket{w_1}{w_3}\}
         \ket{w_2}\bra{w_3}  \nonumber \\ 
    &&\quad\  -\{\braket{w_3}{w_1}-\braket{w_3}{w_2}\braket{w_2}{w_1}\}
        \ket{w_3}\bra{w_1}
       -\{\braket{w_3}{w_2}-\braket{w_3}{w_1}\braket{w_1}{w_2}\}
          \ket{w_3}\bra{w_2}  \nonumber \\ 
    &&\quad\  + \{1-\szetta{\braket{w_1}{w_2}}\}\ket{w_3}\bra{w_3}\  
     \ {\bf ]},
\end{eqnarray}
where
\begin{eqnarray}
  \mbox{det}M&=&1-\szetta{\braket{w_1}{w_2}}-\szetta{\braket{w_2}{w_3}}
      -\szetta{\braket{w_1}{w_3}} \nonumber \\
    &+&\braket{w_1}{w_2}\braket{w_2}{w_3}\braket{w_3}{w_1}
      +\braket{w_1}{w_3}\braket{w_3}{w_2}\braket{w_2}{w_1}.
    \nonumber 
\end{eqnarray}
Perform the calculations of ${\bf\Omega}$, ${\bf\Omega}^{2}$ and 
${\bf\Omega}^{3}$. \quad Moreover 
\begin{flushleft}
{\bf Problem\ {$\infty$}}\quad For the general case perform the similar 
calculations (if possible).
\end{flushleft}
It seems to the author that the calculations in the general case are 
very hard.

\vspace{10mm}
\section{Further Problem}

In this paper we treated generalized coherent states based on $su(1,1)$ 
(\ref{eq:2-2-28-2})
\[
   \ket{w}=\mbox{exp}(wK_{+}-{\bar w}K_{-})\kett{K}{0}\quad 
   \mbox{for}\quad w\in\fukuso
\]
and calculated the Chern characters for $m=1,\ 2$. 

\par \noindent 
But we didn't treat generalized coherent states based on $su(2)$ 
(\ref{eq:2-2-28-1}) 
\[
   \ket{v}=\mbox{exp}(vJ_{+}-{\bar v}J_{-})\kett{J}{0}\quad 
   \mbox{for}\quad v\in\fukuso\subset{\fukuso}P^{1}\ .
\]
See \cite{FKSF1} or \cite{AP}. 
In this case the parameter space is $S^2 \cong {\fukuso}P^{1}$ for $m=1$ 
(a compact manifold). 
To calculate the Chern characters for $m=1,\ 2$ is very interesting 
problem. We leave them to keen graduate students. 

\par \noindent  
It seems to the author that this caluculation will be deeply related to 
the recent non--commutative field theory, see for example 
\cite{SV} or \cite{BDLMO} and their references.

\vspace{10mm}
\section{Discussion}

We have calculated Chern characters for pull--back bundles on ${\cal D}_1$  
and ${\cal D}_2$ which are based on generalized coherent states and 
suggested a relation to some non--commutative field theory (differential 
geometry).
 
\par \noindent 
On the other hand 
one of aims of this paper and \cite{KF9} is to apply the results to 
Quantum Information Theory including Quantum Computation or Quantum 
Cryptgraphy. The study is under progress. 
In the forthcoming paper we would like to discuss this point.

\vspace{1cm}
%
%%%%%%%%%%%%
%References%
%%%%%%%%%%%%


\begin{thebibliography}{99}
%
\bibitem{KS}J. R. Klauder and Bo-S. Skagerstam (Eds) : %1
\newblock Coherent States,
\newblock World Scientific, Singapore, 1985.
%
\bibitem{MW} L. Mandel and E. Wolf : 
\newblock Optical Coherence and Quantum Optics, 
\newblock Cambridge University Press, 1995. 
%
\bibitem{AP}A. Perelomov : %
\newblock Generalized Coherent States and Their Applications,
\newblock Springer--Verlag, 1986.
%
\bibitem{FKSF1}K. Funahashi, T. Kashiwa, S. Sakoda and K. Fujii : %3
\newblock Coherent states, path integral, and semiclassical approximation,
\newblock  J. Math. Phys., 36(1995), 3232.
%
\bibitem{FKSF2}K. Funahashi, T. Kashiwa, S. Sakoda and K. Fujii : %4
\newblock Exactness in the Wentzel-Kramers-Brillouin approximation for 
some homogeneous spaces,
\newblock J. Math. Phys., 36(1995), 4590.
%
\bibitem{FKS}K. Fujii, T. Kashiwa, S. Sakoda :%5
\newblock Coherent states over Grassmann manifolds and the WKB exactness
in path integral,
\newblock J. Math. Phys., 37(1996), 567.
%
\bibitem{KF9} K. Fujii : %13
\newblock Geometry of Coherent States : Some Examples of Calculations 
of Chern Characters, 
\newblock hep--ph/0108219. 
%
\bibitem{MN}M. Nakahara : %14
\newblock Geometry, Topology and Physics,
\newblock IOP Publishing Ltd, 1990.
%
\bibitem{LPS}H-K. Lo, S. Popescu and T. Spiller (eds) : 
\newblock Introduction to Quantum Computation and Information, 
\newblock 1998, World Scientific. 
%
\bibitem{AH} A. Hosoya : 
\newblock Lectures on Quantum Computation (in Japanese), 
\newblock 1999, Science Company (in Japan).
%
\bibitem{KF1} K. Fujii : %1
\newblock Introduction to Grassmann Manifolds and Quantum Computation, 
\newblock quant-ph/0103011.
%
\bibitem{KF2} K. Fujii : %12
\newblock Note on Coherent States and Adiabatic Connections, Curvatures,
\newblock J. Math. Phys.,  
\newblock 41(2000), 4406, 
\newblock quant-ph/9910069. 
%
\bibitem{KF3} K. Fujii : %13
\newblock Mathematical Foundations of Holonomic Quantum Computer,
\newblock to appear in Rept. Math. Phys., 
\newblock quant-ph/0004102.
%
\bibitem{KF4} K. Fujii : %13
\newblock More on Optical Holonomic Quantum Computer,
\newblock quant-ph/0005129.
%
\bibitem{KF5} K. Fujii : %13
\newblock Mathematical Foundations of Holonomic Quantum Computer II,
\newblock quant-ph/0101102.
%
\bibitem{KF6} K. Fujii : %13
\newblock From Geometry to Quantum Computation,
\newblock quant-ph/0107128.
%
\bibitem{GLM}G. M. D'Ariano, L. Maccone and M. G. A. Paris : %6
\newblock Quorum of observables for universal quantum estimation,
\newblock quant-ph/0006006.
%
\bibitem{MP}M. G. A. Paris : %6
\newblock Entanglement and visibility at the output of a Mach--Zehnder 
interferometer, 
\newblock quant-ph/9811078.
%
\bibitem{KB}K. Banaszek : % 
\newblock Optical receiver for quantum cryptography with two coherent 
states, 
\newblock quant-ph/9901067.
%
\bibitem{BW}K. Banaszek and K. Wodkiewicz : %7 
\newblock Direct Probing of Quantum Phase Space by Photon Counting, 
\newblock atom--ph/9603003. 
%
\bibitem{SV}M. Spradlin and A. Volovich : %
\newblock Noncommutative solitons on Kahler manifolds, 
\newblock hep--th/0106180.
%
\bibitem{BDLMO}A. P. Balachandran, B. P. Dolan, J. Lee, X. Martin and 
D. O'Conner : %
\newblock Fuzzy Complex Projective Spaces and their Star--products, 
\newblock hep--th/0107099. 
%
\bibitem{KF7}K. Fujii : %1
\newblock Basic Properties of Coherent and Generalized Coherent 
Operators Revisited,
\newblock Mod. Phys. Lett. A, 
\newblock 16(2001), 1277, 
\newblock quant-ph/0009012.
%
\bibitem{KF8}K. Fujii : %1
\newblock Note on Extended Coherent Operators and Some Basic Properties,  
\newblock quant-ph/0009116.
%
\bibitem{FS}K. Fujii and T. Suzuki : %17
\newblock A Universal Disentangling Formula for Coherent States of Perelomov's
 Type,
\newblock hep-th/9907049.
%
\bibitem{BGi}A. O. Barut and L. Girardello : 
\newblock New ``coherent'' states associated with noncompact groups, 
\newblock Commun. Math. Phys., 
\newblock 21(1971), 222. 
%
\bibitem{FF1}K. Fujii and K. Funahashi :
\newblock Extension of the Barut--Girardello coherent state and path 
integral 
\newblock J. Math. Phys.,
\newblock 38(1997), 4422,
\newblock quant--ph/9708011.
%
\bibitem{FF2}K. Fujii and K. Funahashi :
\newblock Extension of the Barut--Girardello coherent state and path 
integral II, 
\newblock quant-ph/9708041. 
%
\bibitem{Tri}D. A. Trifonov ; 
\newblock Barut--Girardello coherent states for u(p,q) and sp(N,R)
 and their macroscopic superpositions, 
\newblock J. Phys. A , 
\newblock 31(1998), 5673, 
\newblock quant-ph/9711066. 
% 
\end{thebibliography}
\end{document}